\documentclass[preprint,showpacs,showkeys,preprintnumbers,amsmath,amssymb]{revtex4}

\usepackage{graphicx}
\usepackage{dcolumn}
\usepackage{bm}
\usepackage[latin1]{inputenc}

\def\e{\mathrm{e}}

\begin{document}

\author{R. Rossi Jr.}
\email{romeu.rossi@ufv.br}
\affiliation{Universidade Federal de Viçosa, Campus Florestal,
C.P 35690-000 - Florestal, MG - Brasil}

\title{Quantum Bang-Bang control of entangle states}

\begin{abstract}
The effect of quantum ``Bang-Bang" control on entangled states is studied. A system of two initially entangled qubits interacting with a bosonic environment is considered. The interaction induces a loss of the initial entanglement of the two qubits and for specific initial states it causes ``entanglement sudden death". A pulsed control of both qubits leads to the preservation of the entanglement. It is also shown that a single pulse performed after the sudden death time induces an entanglement revival in the two qubits system.
\end{abstract}
\pacs{03.75.Gg, 03.67.Pp}

\maketitle

Entanglement is an exclusive feature of quantum mechanics and represents one of the most counterintuitive phenomenon
predicted by the theory. In the literature entanglement has been extensively studied in the field of quantum foundations and in technological field. Entanglement is an essential ingredient for quantum information and quantum computation, which is known to be extremely more powerful than classical computation \cite{art1}. One of the problems for the implementation of quantum computation is the interaction of the physical system of interest (which would be responsible for such implementation) with the environment and its implications on the entanglement dynamics of the system. Recently, the possibility for the entanglement to vanish in finite time, while coherences decrease asymptotically has been extensively studied\cite{art2,art3,art4}. The so call entanglement sudden death depends on the hamiltonian that governs the dynamics and on the initial state.

In this scenario, the development of strategies to control and protect quantum states against the effects of the interactions with the environment is an essential ingredient for the implementation of quantum information technology. Much attention has been given to this matter, and a considerable number of strategies to prevent decoherence and to preserve entanglement were developed. Some examples of those strategies are: Quantum Zeno Effect (QZE) \cite{art5,art6,art7}, Super Zeno Effect \cite{art8}, strong continuous coupling \cite{art9,art10}.

In ref. \cite{art11} the authors show a strategy to suppress decoherence on a particular spin-boson system. The strategy consists on applications of a time varying control field that acts on the dynamics of the system, reducing unwanted effects of the interaction between environment and system of interest. The strategy was termed quantum ``bang-bang" control. Latter, several applications for this strategy were proposed, in order to give an idea let us quote a set of examples. In Ref.\cite{art11a} the authors studied the effects of pulsed control on a qubit coupled to a quantum critical spin bath. An investigation on the influence of the bang-bang pulses on the dynamics of quantum discord, entanglement, quantum mutual information and classical correlation in a cavity QED system was presented in Ref.\cite{art11b}. In Ref.\cite{art11c} the authors found a sudden transition between classical and quantum decoherence by choosing certain initial states in a two spin system, and propose a scheme to prolong the transition time of the quantum discord by applying the bang-bang pulses. Experimental realizations of the quantum ``bang-bang" control have also been reported \cite{art11d, art11e, art11f}.

In the present work, the quantum ``bang-bang" procedure is implemented to protect entangled states. A system composed by two initially entangled quits interacting with a bosonic environment is considered. The interaction with the environment induces loss of the initial entanglement, and for specific initial entangled states it leads to entanglement sudden death. Performing spin flips (characteristic procedure of quantum bang-bang control) on both qubits we inhibit the loss of entanglement. In the limit of continuous spin flips the entanglement loss is completely washed out. It is also shown that, even after entanglement sudden death, applications of the spin flips can induce a revival of the entanglement.

\section{The system}

Let us consider a system composed by two qubits ($S_{A}$ and $S_{B}$) coupled to a single reservoir. The Hamiltonian of the system is given by

\begin{eqnarray}
H&=&H_{S}+H_{M}+H_{int} \\
H_{S}&=&\hbar\frac{\omega_{0}}{2}\left(\sigma^{(1)}_{z}+\sigma^{(2)}_{z}\right) \notag\\
H_{R}&=&\hbar\sum_{k}\omega_{k}a^{\dagger}_{k}a_{k} \notag\\
H_{int}&=&\hbar\sum_{k}g_{k}\left( \sigma^{(1)}_{z}+\sigma^{(2)}_{z}\right)\left(a^{\dagger}_{k}+a_{k}\right), \notag
\end{eqnarray}
where $\omega_{0}$ is the frequency related to the quantum transition on subsystem $S_{A}$ and $S_{B}$, $\omega_{k}$ are the modes frequencies, $g_{k}$ are the coupling constants and $a_{k}^{\dagger}$ ($a$) is the bosonic creation (annihilation) operator. The interaction Hamiltonian $H_{int}$ is the generalization for two qubits of the spin-boson model \cite{art13}.

In the interaction picture the evolution is governed by the Hamiltonian

\begin{equation}
\tilde{H}(t)=\tilde{H}_{int}(t)=\hbar\left( \sigma^{(1)}_{z}+\sigma^{(2)}_{z}\right)\sum_{k}\left(g_{k}a^{\dagger}_{k}\e^{i\omega_{k}t} + g^{\ast}_{k}a_{k}\e^{-i\omega_{k}t}\right).
\end{equation}

The time evolution operator can be written as:

\begin{equation}
\tilde{U}(t)=\exp\left[\frac{\left( \sigma^{(1)}_{z}+\sigma^{(2)}_{z}\right)}{2}\sum_{k}\left(\alpha_{k} a^{\dagger}_{k} - \alpha^{*}_{k}a_{k}\right)\right],\label{evo}
\end{equation}
where $\displaystyle \alpha_{k}=2g_{k}\frac{1-e^{i\omega_{k}t}}{\omega_{k}}$.

Let us consider that the global system, composed by the two qubits subsystem $S_{A}$-$S_{B}$  ($S$) and the reservoir ($R$), is in a initial state

\begin{equation}
\rho(0)=\rho_{S}\otimes\rho_{R},
\end{equation}
where $\displaystyle\rho_{R}=\prod_{k}\rho_{R,k}(T)=\prod_{k}\left(1-\e^{\beta\hbar\omega_{k}}\right)\e^{-\beta H_{R}}$ represents the reservoir in a thermal equilibrium state at temperature $T$ and $\beta=\frac{1}{k_{B}T}$, with $k_{B}$ the Boltzmann constant. For simplicity, we choose units such that $\hbar=k_{B}=1$.

The evolution of each matrix element of $\rho_{S}$ in the basis $\left\{|00\rangle, |01\rangle, |10\rangle, |11\rangle\right\}$ is given by:

\begin{equation}
\rho_{ij,kl}=\langle ij|Tr_{R}\left(\tilde{U}(t)\rho(0)\tilde{U}^{-1}(t)\right)|kl\rangle,
\end{equation}
where $i;j;k;l=0,1$. Notice that the diagonal elements do not evolve in time.

Let us consider $\rho_{S}(0)$ in a maximal entangled state
\begin{equation}
\rho^{(1)}_{S}(0)=\frac{1}{2}\left(|0,0\rangle+|1,1\rangle\right)\left(\langle0,0|+\langle 1,1|\right).\label{state1}
\end{equation}

The time evolution gives,
\begin{equation}
\rho^{(1)}_{S}(t)=\left(\rho^{(1)}_{00,00}(t)|0,0\rangle\langle0,0|+\rho^{(1)}_{11,11}(t)|1,1\rangle\langle1,1|+\rho^{(1)}_{00,11}(t)|0,0\rangle\langle1,1|+\rho^{(1)}_{11,00}(t)|1,1\rangle\langle0,0|\right),\notag
\end{equation}
where
\begin{eqnarray}
\rho^{(1)}_{00,00}(t)&=&\rho^{(1)}_{11,11}(t)=\frac{1}{2},\\
\rho^{(1)}_{11,00}(t)&=&\frac{1}{2}Tr_{R}\left[\exp\left(2\sum_{k}\left(\alpha_{k} a^{\dagger}_{k} - \alpha^{*}_{k}a_{k}\right)\right)\rho_{R}\right],\\
\rho^{(1)}_{00,11}(t)&=&\frac{1}{2}Tr_{R}\left[\exp\left(-2\sum_{k}\left(\alpha_{k} a^{\dagger}_{k} - \alpha^{*}_{k}a_{k}\right)\right)\rho_{R}\right].
\end{eqnarray}

Following the calculation of $\rho^{(1)}_{11,00}(t)$ we can write,
\begin{equation}
Tr_{R}\left[\exp\left(2\sum_{k}\left(\alpha_{k} a^{\dagger}_{k} - \alpha^{*}_{k}a_{k}\right)\right)\rho_{R}\right]=\prod_{k}Tr_{R}\left[\rho_{R,k}(t)D(\alpha_{k})\right],\label{eq}
\end{equation}
where $D(\alpha_{k})=\exp[\alpha_{k} a^{\dagger}_{k} - \alpha^{*}_{k}a_{k}]$ is the harmonic displacement operator. Notice that for each mode on Eq.(\ref{eq}) the quantity $\left[\rho_{R,k}(t)D(\alpha_{k})\right]$ is the symmetric order generating function for a thermal harmonic oscillator \cite{art14}. Therefore, $\rho^{(1)}_{11,00}(t)$ can be written as
\begin{equation}
\rho^{(1)}_{11,00}(t) = \frac{1}{2}\exp\left(\sum_{k}\frac{|\alpha_{k}|^{2}}{2}\coth\left(\frac{\omega_{k}}{2T}\right)\right).\label{eq2}
\end{equation}

To proceed with the time evolution analysis, we take the continuous limit in Eq. (\ref{eq2})

\begin{equation}
\rho^{(1)}_{11,00}(t) = \frac{1}{2}\exp\left\{16\int_{0}^{\infty}d\omega I(\omega)\coth\left(\frac{\omega}{2T}\right)\frac{
1-\cos\left(\frac{\omega}{2T}\right)}{\omega^{2}}\right\},\label{eq3}
\end{equation}
 where $I(\omega)=\sum_{k}\delta(\omega-\omega_{k})|g_{k}|^{2}$ is the spectral density of the bath. For the calculation that follows, let us consider the spectral density with the form

 \begin{equation}
 I(\omega)=\frac{\eta}{4}\omega\e^{\frac{\omega}{\omega_{c}}},
 \end{equation}
where the parameter $\eta > 0$ is the strength of the system-reservoir coupling and $\omega_{c}$ is the cutoff frequency. For more details about the reservoirs characteristics see \cite{art11} and the references therein.

The system-reservoir interaction affect the initial entanglement between $S_{A}$ and $S_{B}$. To investigate such entanglement dynamics we calculate the concurrence \cite{art16} between $S_{A}$ and $S_{B}$ as a function of time for the initial state on Eq.(\ref{state1})

\begin{equation}
C^{(1)}(t)=2\max\left\{0; |\rho^{(1)}_{01,10}(t)|-\sqrt{\rho^{(1)}_{00,00}(t)\rho^{(1)}_{11,11}(t)}; |\rho^{(1)}_{00,11}(t)|-\sqrt{\rho^{(1)}_{10,10}(t)\rho^{(1)}_{01,01}(t)}\right\}=2|\rho^{(1)}_{00,11}(t)|.
\end{equation}

If we consider the initial state

\begin{equation}
\rho^{(2)}_{A,B}(0)=0.3\left(|0,0\rangle+|1,1\rangle\right)\left(\langle 0,0|+\langle 1,1|\right)+0.2\left(|1,0\rangle\langle 1,0|+|0,1\rangle\langle 0,1|\right),\label{state2}
\end{equation}
for subsystem $S_{A}$ and $S_{B}$, the concurrence is given by:

\begin{equation}
C^{(2)}(t)=2\max\left\{0; |\rho^{(2)}_{11,00}(t)|-\sqrt{\rho^{(2)}_{01,01}(t)\rho^{(2)}_{10,10}(t)}\right\}.
\end{equation}

The loss of entanglement on subsystem $S_{A}$-$S_{B}$ induced by the interaction with the reservoir is illustrated in Fig.1 for the initial state (\ref{state1}) and in Fig.2 for the initial state (\ref{state2}). In Fig.1 the entanglement decrease asymptotically while in Fig.2 the evolution shows the entanglement sudden death. The parameters chosen in both figures corresponds to the high-temperature reservoir \cite{art11}.

\begin{figure}[h]
\centering
  \includegraphics[scale=0.5]{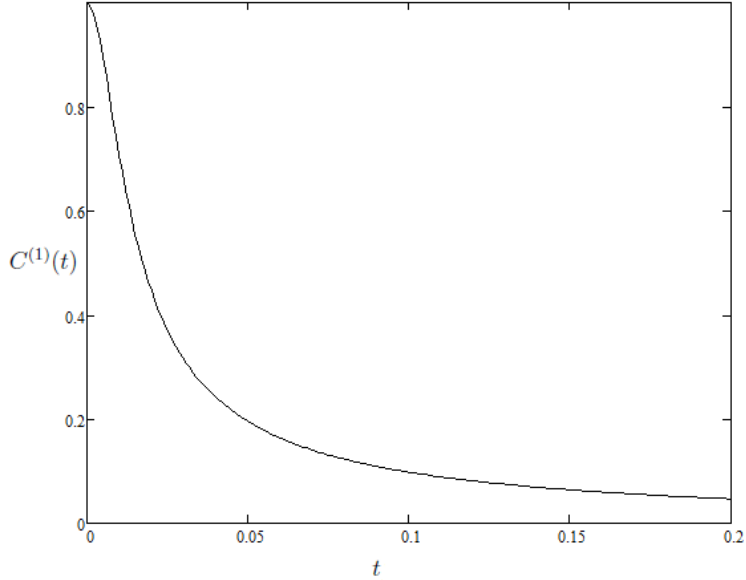}\\
  \caption{Concurrence $C^{(1)}$ as function of $t$. Time is in units of $T^{-1}$ and $\omega_{c}$ in units of $T$. The parameter $\eta=0.25$ and $\omega_{c}/T=100$.}
\end{figure}

\begin{figure}[h]
\centering
  \includegraphics[scale=0.5]{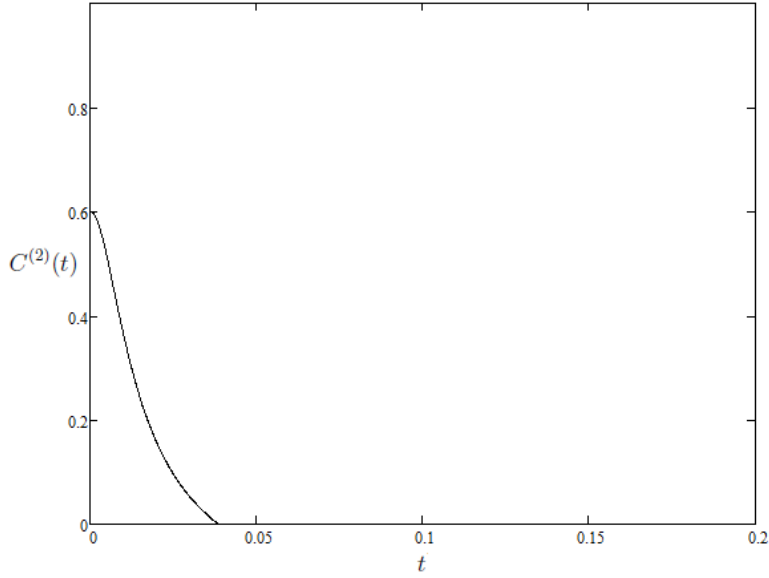}\\
  \caption{Concurrence $C^{(1)}$ as function of $t$. Time is in units of $T^{-1}$ and $\omega_{c}$ in units of $T$. The parameter $\eta=0.25$ and $\omega_{c}/T=100$.}
\end{figure}

\section{Entanglement Protection}

In this section a quantum ``bang-bang" control of the entanglement on system the $S_{A}$-$S_{B}$ is shown. Such control is against the loss induced by interaction with the environment. The quantum ``bang-bang" control is based on ative external control performed by a sequence of pulses acting on a spin system. In the quantum ``bang-bang" control of the entanglement on the subsystem $S_{A}$-$S_{B}$, we consider a sequence of pulses acting (at the same time) on the subsystem $S_{A}$ and on the subsystem $S_{B}$. Each pulse can be represented by an unitary operator. The hamiltonian that governs the time evolution of the global system submitted to a sequence of pulses in Schrodinger picture is given by

\begin{equation}
H(t)=H_{int} + H^{(A)}_{p}(t)+H^{(B)}_{p}(t),
\end{equation}
where
\begin{equation}
H^{(K)}_{p}(t)=\sum_{n=1}^{N}V^{n}(t)\left\{\cos\left[\omega_{0}\left(t-t_{p}^{(n)}\right)\right]\sigma_{x}^{(K)}+
\sin\left[\omega_{0}\left(t-t_{p}^{(n)}\right)\right]\sigma_{y}^{(K)}\right\},
\end{equation}
represents a monochromatic alternating magnetic field at resonance, applied on the subsystem $S_{K}$, with $K=A,B$. The function $V^{n}(t)$ is a constant $V$ for the entire duration of each pulse and is zero elsewhere. The number of pulses in the sequence is $N$, each of duration $\tau_{p}$ and applied at instants $t_{p}^{(n)}=t_{0}+n\Delta t$. The interval between pulses is $\Delta t$.

To perform the quantum ``bang-bang" control we consider pulses satisfying the condition $2 V\tau_{p}=\pi$, and suppose that $V$
is large enough to yield almost instantaneous pulses, i.e., $\tau_{p}\rightarrow 0$ and $V\rightarrow\infty$ such that  $V\tau_{p}=\frac{\pi}{2}$. The effect of each pulse is a spin flip on the subsystem $S_{A}$ and on the subsystem $S_{B}$.

In the interaction picture the unitary evolution operator for the $j$-th pulse can be written as
\begin{equation}
\tilde{U}_{p_{j}}=\tilde{U}_{p_{j}}^{A}\otimes\tilde{U}_{p_{j}}^{B}=\exp\left\{i\omega_{0}\sigma_{z}^{A}t_{p}^{(j)}\right\}\sigma_{x}^{(A)}\otimes
\exp\left\{i\omega_{0}\sigma_{z}^{B}t_{p}^{(j)}\right\}\sigma_{x}^{(B)}.
\end{equation}

For details on the calculation of $U_{p_{j}}^{A}$ and $U_{p_{j}}^{B}$ see Ref.\cite{art11}.

\subsection{Single pulse}

To illustrate the effect produced by the pulses on the entanglement dynamics of the two qubits system, we investigate a single pulse that divides the interaction between $S_{A}$-$S_{B}$ and $R$ in two steps. Let us consider the evolution of the initial state $\rho^{(1)}_{S}(0)$ (Eq. (\ref{state1})) given by:

\begin{equation}
\rho^{(1)}_{S}(t)=Tr_{R}\left\{\tilde{U}(t-t^{(1)}_{p})\left(\tilde{U}_{p_{1}}\otimes I_{(R)}\right)\tilde{U}(t^{(1)}_{p})\rho^{(1)}_{S}(0)\otimes\rho_{R}\tilde{U}^{\dagger}(t-t^{(1)}_{p})\left(\tilde{U}_{p_{1}}\otimes I_{(R)}\right)^{\dagger}\tilde{U}^{\dagger}(t^{(1)}_{p})\right\},
\end{equation}
where $I_{(R)}$ is the identity operator on the subsystem $R$. In the first step of the evolution ($0\leq t<t^{(1)}_{p}$), the spin system interacts with the reservoir. At the time $t=t^{(1)}_{p}$ the spin system undergo a pulse. In the second step of the evolution ($t>t^{(1)}_{p}$) the spin system interacts again with the reservoir. The time evolution is given by

\begin{equation}
\rho^{(1)}_{S}(t)=\left(\rho^{(1)}_{00,00}(t)|0,0\rangle\langle0,0|+\rho^{(1)}_{11,11}(t)|1,1\rangle\langle1,1|+\rho^{(1)}_{00,11}(t)|0,0\rangle\langle1,1|+\rho^{(1)}_{11,00}(t)|1,1\rangle\langle0,0|\right),\notag
\end{equation}
where
\begin{eqnarray}
\rho^{(1)}_{00,00}(t)&=&\rho^{(1)}_{11,11}(t)=\frac{1}{2},\\
\rho^{(1)}_{11,00}(t)&=&\frac{1}{2}Tr_{R}\left\{e^{A}\rho_{R}\right\},\notag\\
\rho^{(1)}_{00,11}(t)&=&\frac{1}{2}Tr_{R}\left\{e^{-A}\rho_{R}\right\},
\end{eqnarray}
and
\begin{equation}
A=\left\{2\sum_{k}\left[e^{i\omega_{k}t_{p}^{(1)}}\alpha_{k}(t-t_{p}^{(1)})
-\alpha_{k}(t_{p}^{(1)})\right] a^{\dagger}_{k} - \left[e^{-i\omega_{k}t_{p}^{(1)}}\alpha^{*}_{k}(t-t_{p}^{(1)})
-\alpha^{*}_{k}(t_{p}^{(1)})\right]a_{k}\right\}.
\end{equation}

 The spin flip induces a change in the entanglement evolution, part of the entanglement lost on the first step is recovered in the second step. To make it clear we calculate the concurrence $C^{(1)}(t)=2|\rho^{(1)}_{11,00}(t)|$ and show the effect of a single pulse in a graphic (Fig.3). In this calculation we apply the same method used in the first section to calculate the matrix element in Eq.(\ref{eq2}) and (\ref{eq3}).

\begin{equation}
C^{(1)}(t)=2|\rho^{(1)}_{11,00}(t)|=\exp\left\{4\eta\int_{0}^{\infty}e^{-\omega/\omega_{c}}\coth\left(\frac{\omega}{2T}\right)B(\omega)\right\},
\end{equation}
where

\begin{equation}
B(\omega)=\frac{2-\cos\left[\omega\left(t-t_{p}^{(1)}\right)\right] -\cos\left(\omega t_{p}^{(1)}\right)-2Re\left[e^{i\omega t^{(1)}_{p}}\left(1-e^{i\omega\left(t-t_{p}^{(1)}\right)}\right)\left(1-e^{-i\omega t_{p}^{(1)}}\right)\right]}{\omega}.\label{b}
\end{equation}

The graphic in Fig.3 clearly show the change on the entanglement dynamics induced by a single pulse performed at the time $t=t^{(1)}_{p}$. In the time interval $0\leq t<t^{(1)}_{p}$ the subsystem $S_{A}$-$S_{B}$ interacts with the reservoir $R$ and the entanglement of the spin system decrease as it was shown in Fig.1. At the time $t=t^{(1)}_{p}$ the qubit system undergo a pulse (the spin flip described in the previous section). For $t>t^{(1)}_{p}$ the spin system interacts again with the reservoir, but in this second step of the interaction the entanglement increases for a finite time interval. Part of the initial spin system entanglement that was lost on the first step of the interaction is recovered and the concurrence reaches a local maximum. After the entanglement decreases again.

\begin{figure}[h]
\centering
  \includegraphics[scale=0.5]{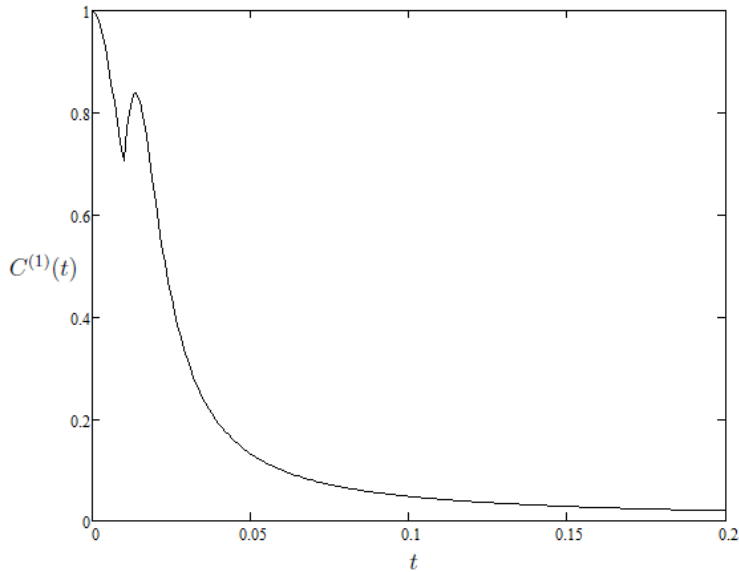}\\
  \caption{Concurrence $C^{(1)}$ as function of $t$. Time is in units of $T^{-1}$ and $\omega_{c}$ in units of $T$. The parameter $\eta=0.25$ and $\omega_{c}/T=100$. At the time $t^{(1)}_{p}=0.01$ the qubit system undergo a pulse.}
\end{figure}

\subsection{Revival of entanglement}

Let us now consider the initial sate $\rho^{(2)}_{S}$ of Eq.(\ref{state2}). As it is shown in the first section, the entanglement dynamics of $\rho^{(2)}_{S}$ presents entanglement sudden death. In this section it is shown that if a single pulse is performed on the spin system, after entanglement sudden death time, it can induce a revival of entanglement.

The evolution of $\rho^{(2)}_{S}$ with a single pulse performed at $t=t^{(1)}_{p}$ is given by

\begin{equation}
\rho^{(2)}_{S}(t)=Tr\left\{\tilde{U}(t-t^{(1)}_{p})\left(\tilde{U}_{p_{1}}\otimes I_{(R)}\right)\tilde{U}(t^{(1)}_{p})\rho^{(2)}_{S}(0)\otimes\rho_{R}\tilde{U}^{\dagger}(t-t^{(1)}_{p})\left(\tilde{U}_{p_{1}}\otimes I_{(R)}\right)^{\dagger}\tilde{U}^{\dagger}(t^{(1)}_{p})\right\}.
\end{equation}

The entanglement dynamics is given by the concurrence

\begin{equation}
C^{(2)}(t)=2\max\left\{0; |\rho^{(2)}_{11,00}(t)|-\sqrt{\rho^{(2)}_{01,01}(t)\rho^{(2)}_{10,10}(t)}\right\},
\end{equation}
following the calculation in previous sections

\begin{eqnarray}
\rho^{(2)}_{01,01}(t)&=&\rho^{(1)}_{10,10}(t)=0.2,\\
\rho^{(2)}_{11,00}(t)&=&0.3\exp\left\{4\eta\int_{0}^{\infty}e^{-\omega/\omega_{c}}\coth\left(\frac{\omega}{2T}\right)B(\omega)\right\},\notag\\
\rho^{(2)}_{00,11}(t)&=&0.3\exp\left\{4\eta\int_{0}^{\infty}e^{-\omega/\omega_{c}}\coth\left(\frac{\omega}{2T}\right)B(\omega)\right\},\notag
\end{eqnarray}
where $B(\omega)$ is given by Eq.(\ref{b}).

In Fig.4 it is shown the entanglement revival. The spin flip performed after the sudden death time induces the entanglement revival and part of the two qubits system initial entanglement is recovered.

\begin{figure}[h]
\centering
  \includegraphics[scale=0.5]{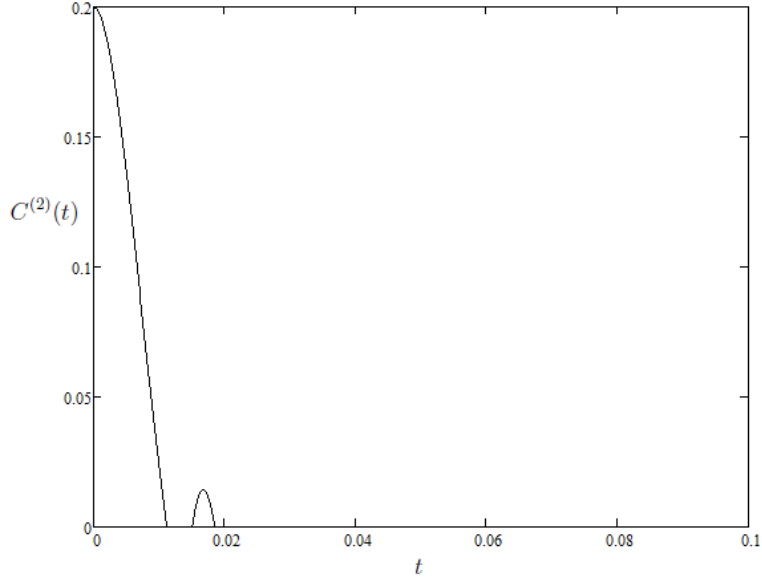}\\
  \caption{Concurrence $C^{(2)}$ as function of $t$. Time is in units of $T^{-1}$ and $\omega_{c}$ in units of $T$. The parameter $\eta=0.25$ and $\omega_{c}/T=100$. At the time $t^{(1)}_{p}=0.0135$ (after sudden death time) the qubit system undergo a pulse.}
\end{figure}

\subsection{A sequence of pulses: Entanglement quantum bang-bang control}

In this section we show the quantum ``bang-bang" control of the spin system entanglement. The evolution that represents the ``bang-bang" control is a sequence of $N$ pulses (where $N$ is an even number) and can be written as

\begin{eqnarray}
\tilde{U}_{N}=\tilde{U}_{p_{N}}\tilde{U}(\Delta t)\ldots\tilde{U}_{p_{n+1}}\tilde{U}(\Delta t)\tilde{U}_{p_{n}}\tilde{U}(\Delta t)\ldots\tilde{U}_{p_{1}}\tilde{U}(\Delta t).
\end{eqnarray}

Let us consider a sequence of two pulses, the $n$-th and $n+1$th. The pulses are performed at time $t^{(n)}_{p}$ and $t^{(n+1)}_{p}$ respectively. The time evolution operator, that represents this sequence, written in the basis $\left\{|00\rangle, |01\rangle, |10\rangle, |11\rangle\right\}$ is given by

\begin{eqnarray}
\tilde{U}_{n,n+1}=\tilde{U}_{p_{n+1}}\tilde{U}(\Delta t)\tilde{U}_{p_{n}}\tilde{U}(\Delta t)&=&e^{-2i\omega_{0}(t^{(n)}_{p}+ t^{(n+1)}_{p})}[|0,1\rangle\langle0,1|+|1,0\rangle\langle1,0|\\
&+&e^{D}|0,0\rangle\langle0,0|+e^{-D}|1,1\rangle\langle0,0|],
\end{eqnarray}
where

\begin{equation}
D=\sum_{k}a_{k}^{\dagger}e^{i\omega_{k}t_{0}}\alpha_{k}(\Delta t)\left(e^{i\omega_{k}\Delta t}-1\right)+
a_{k} e^{-i\omega_{k}t_{0}}\alpha^{*}_{k}(\Delta t)\left(e^{-i\omega_{k}\Delta t}-1\right).
\end{equation}

If the time internal between two pulses goes to zero, $\Delta t \rightarrow 0$, we have $D \rightarrow 0$ and $\tilde{U}_{p_{n+1}}\tilde{U}(\Delta t)\tilde{U}_{p_{n}}\tilde{U}(\Delta t) \rightarrow I$, where $I$ is the identity operator of the global system.

Notice that we can also write $\tilde{U}_{N}$ as

\begin{equation}
\tilde{U}_{N}=\tilde{U}_{N-1,N}\ldots\tilde{U}_{n,n+1}\ldots\tilde{U}_{1,2}.
\end{equation}

When $\Delta t \rightarrow 0$ we have  $\tilde{U}_{N}\rightarrow I$ and consequently, the preservation of the initial state of the global system.

In conclusion, it is shown the quantum ``bang-bang" control of entanglement in a system of two qubits $S_{a}$ and $S_{b}$. The control is performed by a sequence of pulses applied on the subsistem $S_{a}$-$S_{b}$. When the time interval between the pulses goes to zero, the initial entanglement on $S_{a}$-$S_{b}$ is completely preserved. It is also shown that for initial states of $S_{a}$-$S_{b}$ whose evolution presents entanglement sudden death, a single pulse applied after the sudden death time can induce a revival of the entanglement.


\begin{thebibliography}{21}




\bibitem{art1} M. A. Nielsen and  I. L. Chuang, {\it Quantum Computation
and Quantum Information} (Cambridge University Press, Cambridge,
England, 2000).


\bibitem{art2} T. Yu and J.H. Eberly, \prl \textbf{93}, 140404 (2004).



\bibitem{art3} T. Yu and J.H. Eberly, \prl \textbf{97}, 140403 (2006).



\bibitem{art4} Z. Ficek, R. Tanas, \pra \textbf{74}, 024304(2006).



\bibitem{art5} Y. P. Huang and M. G. Moore, \pra \textbf{77}, 062332 (2008).



\bibitem{art6} J. D. Franson, B. C. Jacobs, and T. B. Pittman, \pra \textbf{70}, 062302 (2004).


\bibitem{art7} S. Maniscalco, F. Francica, R. L. Zaffino, N. Lo Gullo and F. Plastina,
\prl \textbf{100}, 090503 (2008).


\bibitem{art8} D. Dhar, L. K. Grover, and S. M. Roy, \prl \textbf{96}, 100405 (2006).


\bibitem{art9} P. Facchi and S. Pascazio, \prl \textbf{89}, 080401
(2002).


\bibitem{art10} A. Beige, D. Braun, B. Tregenna, and P. L. Knight,
\prl \textbf{85}, 1762 (2000).



\bibitem{art11} L. Viola and S. Lloyd, \pra \textbf{58}, 2733 (1998).



\bibitem{art11a} D. Rossini, P. Facchi, R. Fazio, G. Florio, D. A. Lidar, S.Pascazio, F. Plastina and P. Zanardi, \pra \textbf{77},052112 (2008).



\bibitem{art11b} H.-S. Xu and J.-B. Xu, European Physics Letters \textbf{95}, 60003 (2011).



\bibitem{art11c} Da-Wei Luo, Hai-Qing Lin, Jing-Bo Xu, and Dao-Xin Yao, \pra 84, 062112 (2011).



\bibitem{art11d} S. Damodarakurup, M. Lucamarini, G. Di Giuseppe, D. Vitali, and P. Tombesi, \prl \textbf{103}, 040502 (2009).



\bibitem{art11e} J.J.L. Morton, A.M. Tyryshkin, A. Ardavan, S.C. Benjamin, K. Porfyrakis, S. A. Lyon and G. Andrew D. Briggs, Nature Physics \textbf{2}, 40 (2006).




\bibitem{art11f} M. Lucamarini, G. Di Giuseppe, S. Damodarakurup, D. Vitali, and P. Tombesi, \pra \textbf{83}, 032320 (2011).



\bibitem{art13} David P. DiVincenzo \pra \textbf{51}, 1015 (1995).



\bibitem{art14} M. Hillery \emph{et. al}., Phys. Rep. \textbf{106}, 121 (1984).



\bibitem{art15} H. P. Breuer and F. Petruccione, {\it The Theory of Open Quantum System} (Oxford
University Press, 2002).


\bibitem{art16} S. Hill and W. K. Wootters, \prl, \textbf{78}, 5022 (1997).


\end{thebibliography}
\end{document}